\documentclass[8.5pt,twoside,twocolumn]{article}
\pdfoutput=1
\usepackage[super,sort&compress,comma]{natbib} 
\usepackage{mhchem}
\usepackage{times,mathptmx}
\usepackage{sectsty}
\usepackage{balance} 
\usepackage[english]{babel}
\usepackage{graphicx} 
\usepackage{lastpage}
\usepackage[format=plain,justification=raggedright,singlelinecheck=false,font=small,labelfont=bf,labelsep=space]{caption} 
\usepackage[breaklinks,colorlinks,urlcolor=blue,citecolor=blue,linkcolor=blue]{hyperref}
\usepackage{url}
\usepackage{bm}
\usepackage{fancyhdr}
\newcommand{\new}{}
\begin{document}

\thispagestyle{plain}
\fancypagestyle{plain}{
\fancyhead[L]{\includegraphics[height=8pt]{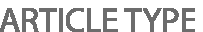}}
\fancyhead[C]{\hspace{-1cm}\includegraphics[height=20pt]{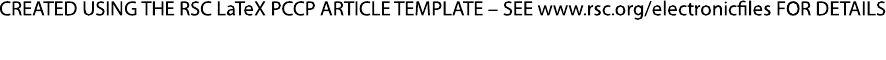}}
\fancyhead[R]{\includegraphics[height=10pt]{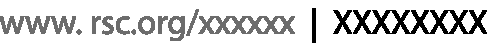}\vspace{-0.2cm}}
\renewcommand{\headrulewidth}{1pt}}
\renewcommand{\thefootnote}{\fnsymbol{footnote}}
\renewcommand\footnoterule{\vspace*{1pt}%
\hrule width 3.4in height 0.4pt \vspace*{5pt}} 
\setcounter{secnumdepth}{5}

\makeatletter 
\def\subsubsection{\@startsection{subsubsection}{3}{10pt}{-1.25ex plus -1ex minus -.1ex}{0ex plus 0ex}{\normalsize\bf}} 
\def\paragraph{\@startsection{paragraph}{4}{10pt}{-1.25ex plus -1ex minus -.1ex}{0ex plus 0ex}{\normalsize\textit}} 
\renewcommand\@biblabel[1]{#1}            
\renewcommand\@makefntext[1]%
{\noindent\makebox[0pt][r]{\@thefnmark\,}#1}
\makeatother 
\renewcommand{\figurename}{\small{Fig.}~}
\sectionfont{\large}
\subsectionfont{\normalsize} 

\fancyfoot{}
\fancyfoot[LO,RE]{\vspace{-7pt}\includegraphics[height=9pt]{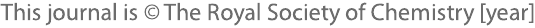}}
\fancyfoot[CO]{\vspace{-7.2pt}\hspace{12.2cm}\includegraphics{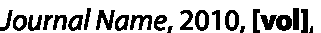}}
\fancyfoot[CE]{\vspace{-7.5pt}\hspace{-13.5cm}\includegraphics{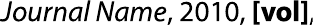}}
\fancyfoot[RO]{\footnotesize{\sffamily{1--\pageref{LastPage} ~\textbar  \hspace{2pt}\thepage}}}
\fancyfoot[LE]{\footnotesize{\sffamily{\thepage~\textbar\hspace{3.45cm} 1--\pageref{LastPage}}}}
\fancyhead{}
\renewcommand{\headrulewidth}{1pt} 
\renewcommand{\footrulewidth}{1pt}
\setlength{\arrayrulewidth}{1pt}
\setlength{\columnsep}{6.5mm}
\setlength\bibsep{1pt}

\twocolumn[
  \begin{@twocolumnfalse}
\noindent\LARGE{\textbf{Interactions of adsorbed CO$_2$ on water ice at low temperatures}}
\vspace{0.6cm}

\noindent\large{\textbf{L.J. Karssemeijer,\textit{$^{a}$} G.A. de Wijs,\textit{$^{b}$} and H.M. Cuppen$^{\ast}$\textit{$^{a}$}} }\vspace{0.5cm}

\noindent\textit{\small{\textbf{Received Xth XXXXXXXXXX 20XX, Accepted Xth XXXXXXXXX 20XX\newline
First published on the web Xth XXXXXXXXXX 200X}}}

\noindent \textbf{\small{DOI: 10.1039/b000000x}}
\vspace{0.6cm}

\noindent \normalsize{We present a computational study into the adsorption properties of CO$_2$ on amorphous and crystalline water surfaces under astrophysically relevant conditions. Water and carbon dioxide are two of the most dominant species in the icy mantles of interstellar dust grains and a thorough understanding of their solid phase interactions at low temperatures is crucial for understanding the structural evolution of the ices due to thermal segregation. In this paper, a new H$_2$O-CO$_2$ interaction potential is proposed and used to model the ballistic deposition of CO$_2$ layers on water ice surfaces, and to study the individual binding sites at low coverages. Contrary to recent experimental results, we do not observe CO$_2$ island formation on any type of water substrate. Additionally, density functional theory calculations are performed to assess the importance of induced electrostatic interactions.}

\vspace{0.5cm}
 \end{@twocolumnfalse}
  ]

\footnotetext{\textit{$^{a}$~Theoretical Chemistry, Institute for Molecules and Materials, Radboud University Nijmegen, Heyendaalseweg 135, 6525 AJ Nijmegen, The Netherlands. E-mail: hcuppen@science.ru.nl} }
\footnotetext{\textit{$^{b}$~Electronic Structure of Materials, Institute for Molecules and Materials, Radboud University Nijmegen, Heyendaalseweg 135, 6525 AJ Nijmegen, The Netherlands. } }

\section{Introduction}
The interactions of molecular species with the surface of water ices are of fundamental importance in interstellar chemistry~\cite{Tielens2010} and in atmospheric science~\cite{Andersson2004}. In the cold and dense regions of the interstellar medium molecules adsorb, diffuse, and react on the surface of icy dust grain mantles, composed mainly of H$_2$O~\cite{Herbst2009a,Tielens2013}. A detailed knowledge of these physicochemical surface processes is needed to accurately model the physics and chemistry of molecular clouds and the composition of the dust grain mantles~\cite{Wakelam2010,Chang2012,Cuppen2013}. 

Laboratory techniques, in particular Temperature Programmed Desorption (TPD), provide a very useful tool to probe the surface-adsorbate interactions under typical molecular cloud conditions and have been used to extract desorption energies for many molecules of astrophysical interest~\cite{Collings2004a,Burke2010,Noble2012a}. However useful, these experiments measure average quantities and lack the resolution to probe the systems on the nanoscale and assess effects of, for example, surface inhomogeneities, which have a big influence on surface reactivity and diffusive properties~\cite{Cuppen2011a,Karssemeijer2014}.
From the theoretical side, atomistic simulation methods can probe specific processes at the molecular scale, and can be used to study reaction mechanisms~\cite{Goumans2009a}, photochemical processes~\cite{Arasa2010}, and thermal kinetics over long timescales~\cite{Karssemeijer2012}. To use these methods, accurate descriptions of the intermolecular interactions in the system are required. Unfortunately, these are not always available because the large simulation cells needed to model the amorphous ice mantles, require very efficient interaction potentials.

In the present paper, we present a computational study into the adsorption properties of CO$_2$ on crystalline and amorphous water ices. Carbon dioxide is an abundant and ubiquitous component of interstellar ices with typical abundances of 0.15 to 0.40 with respect to water~\cite{Pontoppidan2008}. As such, CO$_2$ plays an important role in the interstellar ice chemistry. Furthermore, the infrared spectrum of solid CO$_2$ is very sensitive to the local molecular environment and as such, it can be used to infer information about the structure of the ices and the physical conditions of the molecular cloud.~\cite{Sandford1990a,Gerakines1999,Pontoppidan2008,Ehrenfreund1998}. \citet{Pontoppidan2008} showed that, by decomposing the 15.2~$\mu$m band into five distinct components, the segregation of mixed H$_2$O:CO$_2$ and CO$_2$:CO ices into separated layers or pockets, can be observed. As the temperature in a collapsing molecular cloud increases towards a young protostar, the component associated with pure CO$_2$ is seen to increase, at the expense of the mixed features. Combined with detailed knowledge of the rate of such kinetic processes from experiments or theory, these observations can be used to study the thermal history of molecular clouds. At the moment however, such data is scarcely available.

The segregation of mixed ices (H$_2$O:CO$_2$ and H$_2$O:CO) has been studied by \citet{Oberg2009} using a combination of both infrared spectroscopy and a kinetic Monte Carlo model. For this model however, estimates had to be made regarding the segregation barriers and binding energies, as data from TPD experiments were not sufficient~\cite{Oberg2009}. A key question which remained was whether the H$_2$O-CO$_2$ binding energy is stronger or weaker than the CO$_2$-CO$_2$ interactions. At this moment still, experiments have not given a decisive answer. From recent, submonolayer TPD results by \citet{Noble2012a}, it was argued that CO$_2$ is a non-wetting molecule on water substrates, which will prefer the formation of islands over the formation of a homogeneous surface layer. This claim was partly based on a model fit to the experiments, where at coverages just before one monolayer, the binding energy was weaker than for multilayer coverages. This multilayer binding energy was found to be 195~meV on amorphous solid water (ASW). At the very lowest coverages though, the binding energy was even stronger, namely 202~meV. Only in the intermediate regime did the binding energy fall below multilayer value. Furthermore, the TPD spectra of CO$_2$ desorption from ASW showed the onset of multilayer features already before monolayer coverage was reached. Interestingly though, this feature was only observed on ASW, and not on crystalline ice. In another recent TPD study, these features are not observed~\cite{Edridge2010,Edridge2013}. Also in this, last mentioned, experiment, the multilayer binding energy of CO$_2$ on ASW was found to be stronger than the monolayer value ($264\pm15$ versus $181\pm78$~meV), but these multilayer values are very different from the first experiment and the monolayer values are found to be broadly distributed. In the multilayer regime, several other studies have been performed on various substrates, but the results are varying~\cite{Sandford1990a,Galvez2007,Andersson2004,Ulbricht2006}.

Based on the above, the adsorption behavior of CO$_2$ on water ice is complex, mainly because the substrate-adsorbate interaction energy is comparable to the adsorbate-adsorbate interactions. To understand a complicated process, like the segregation of mixed interstellar ices however, an accurate description of the system is needed. We have therefore studied the adsorption behavior of CO$_2$ on two condensed phases of water, hexagonal ice and amorphous solid water. The key question which is addressed is whether or not CO$_2$ shows wetting behavior on these two different substrates. This behavior will already be indicative of the mechanisms underlying bulk segregation, which should be an interesting subject for future investigations. These can be performed, if an accurate and computationally efficient interaction model is available.

In this work, we present a new H$_2$O-CO$_2$ pair potential (PP) model, based on accurate \textit{ab-initio} calculations of the gas-phase complex. The binding energy of this complex is about twice as large as that of the the CO$_2$ dimer (124~meV\cite{Makarewicz2010} versus 60~meV~\cite{Bukowski1999}), so based purely on gas-phase data, one would expect CO$_2$ to be a fully wetting molecule on a water surface. Of course, the solid water substrate presents a totally different system and adsorption behavior can be heavily influenced by steric effects and non-additive interactions~\cite{Elrod1994}. The steric effects are captured with the PP model, but non-additive interactions are inherently not included. To investigate these, we perform additional density function theory (DFT) calculations and evaluate the induced dipole-dipole interactions. The computational details are given in Section~\ref{sec:compdetails}, followed by a discussion of the results in Section~\ref{sec:results}. Conclusions are drawn in the final Section.
\begin{table}[h]
\caption{Parameters for the H$_2$O-CO$_2$ Buckingham potential.}
\centering
\begin{tabular}{cccc}
\hline
Interaction & $A_{ij}$ (eV) & $B_{ij}$ (eV\AA$^{-1}$) & $C_{ij}$ (eV\AA$^{6}$) \\\hline
H C & 80.71 & 3.006 & 7.395 \\
H O & 40.45 & 3.521 & 8.540$\times10^{-2}$ \\
O C & 47.97 & 2.480 & 19.87 \\
O O & 5496  & 3.927 & 30.98 \\\hline
\end{tabular}
\label{tab:vh2oco2}
\end{table} 

\section{Computational Details}
\label{sec:compdetails}
Most of the results presented in this paper are computed using the PP model, which we describe in Section~\ref{sec:pairpotentials}. Then, the details of the DFT calculations are outlined in Section~\ref{sec:dfttheory}. To allow comparison to experimental values, all binding energies presented in this work have been corrected for the zero point energy contribution, following the procedure described in \citet{Karssemeijer2014}. \new{The corrections applied to the DFT results are the same as the ones applied on the PP binding energies}.

\subsection{Pair interactions}
\label{sec:pairpotentials}
In the pair interaction model, forces and energies are derived from pairwise interactions between each molecule-molecule pair in the system. All the molecules are fully flexible, within their own intramolecular potential. Because we are considering two molecular species, three interaction potentials are needed. The H$_2$O-H$_2$O interactions are modeled with the semi-empirical TIP4P/2005f~\cite{Gonzalez2011} potential. This is a flexible version of the TIP4P/2005f potential which was fitted specifically to describe the condensed phases of water~\cite{Abascal2005}. For the CO$_2$-CO$_2$ interactions, we use the EPM potential by~\citet{Harris1995}. We chose this potential because it reproduced the CO$_2$ dimer energies of the accurate SAPT-a potential from~\citet{Bukowski1999} with a satisfactory accuracy. Because the EPM potential contains fixed C-O bond length, the harmonic intramolecular potential from~\citet{Zhu1989} is used. 

A satisfactory H$_2$O-CO$_2$ potential was not available from literature. Therefore, we fitted a model potential to \textit{ab-initio} energy calculations on a set of 316 configurations of the two molecule complex, at 30 distinct angular orientations and monomer separations between 1.5 and 20~\AA. These calculations were done at the CCSD(T) level with the aug-cc-pVTZ basis set using the \textsc{gaussian03}~\cite{Gaussian03} and \textsc{molpro2010}~\cite{MOLPRO} packages. Corrections were made for the basis set superposition error with the Boys-Bernardi counterpoise scheme~\cite{Boys1970}. \new{We used \textit{ab-initio} calculations to fit our potential because we want to reproduce the molecular geometries and energetics on the single-molecule scale as accurately as possible in order to evaluate individual binding sites of adsorbed CO$_2$ molecules. For this purpose, we think using \textit{ab-initio} data is a better approach than relying on bulk experimental data, as is often done to fit effective, empirical, pair potentials.} The proposed H$_2$O-CO$_2$ potential contains a point-charge based electrostatic term and a site-site term accounting for the van der Waals (vdW) interactions. For the electrostatics, the charges on the H$_2$O and CO$_2$ molecules are identical to those in the H$_2$O-H$_2$O and CO$_2$-CO$_2$ potentials. The vdW interactions are modeled by a Buckingham potential between each intermolecular atomic pair in the system:
\begin{equation}
V = \sum_{i\in \textrm{H}_2\textrm{O}} \sum_{j\in \textrm{CO}_2} A_{ij}\exp\left(-B_{ij}r_{ij}\right) + \frac{C_{ij}}{r_{ij}^6},
\end{equation}
where $r_{ij}$ is the distance from atom $i$ to atom $j$. The parameters of this potential are given in Table~\ref{tab:vh2oco2}. All pair interactions are smoothly cut off between \new{9 and 10~\AA, based on the molecular center of mass distance, $r_{\textrm{com}}$, using the switching function $f(x) = (2x-3)x^2+1$ with $x\in (0,1)$, so $x=r_{\textrm{com}}-9$\AA.}

\subsection{Density Functional Calculations}
\label{sec:dfttheory}
For our purpose, it is important that the DFT calculations reproduce both the structure of the ice and the interactions with carbon dioxide correctly. The PBE exchange and correlation functional~\cite{Perdew1996} is widely used to study hexagonal ice~\cite{Hamann1997,Feibelman2008,Raza2011,Santra2013} but does not describe vdW interactions accurately.  For our purpose however, a correct treatment of these interactions is important, because the H$_2$O-CO$_2$ interaction is dominated by dispersion interactions. The semi-empirical DFT-D2 method~\cite{Grimme2006} can correct for this at negligible computational cost, but the additional attractive energy leads to significant overbinding in the ice~\cite{Raza2011}, in combination with the PBE functional.  We therefore used the van der Waals inclusive optPBE-vdW exchange and correlation functional~\cite{Klimes2010} which treats the vdW interactions within the vdW-DF approach~\cite{Dion2004}. The functional was recently shown to give a good description of hexagonal ice~\cite{Santra2013}.

All DFT calculations were performed with a plane wave basis set using the Vienna \textit{Ab-initio} Simulation Package (VASP)~\cite{Kresse1996,Kresse1996a} using the projector augmented-wave (PAW) method~\cite{Blochl1994,Kresse1999}, in which the optPBE-vdW functional is included~\cite{Klimes2011}. Based on convergence tests, we used a $\Gamma$-centered Monkhorst-Pack $\bm{k}$-point mesh~\cite{Monkhorst1976} with a spacing less than 0.04~\AA$^{-1}$ along each reciprocal lattice vector and a plane wave energy cutoff of 600~eV. The standard PAW data sets provided by VASP are used and the atomic structures are considered to be converged when the force on any atom in the system is less than 0.05~eV~\AA$^{-1}$.

To evaluate the induced electrostatics of the systems, the electronic dipole moments of individual molecules are extracted from the DFT calculations by transforming the delocalized Bloch orbitals into Maximally Localized Wannier Functions (MLWFs)~\cite{Mostofi2008,Marzari2012}. This procedure provides a set of Wannier function centers, which correspond to each MLWF. Based on their positions, these centers can be uniquely assigned to individual molecules in the system and by placing electronic charges at the centers, the dipole moment of each molecule is readily evaluated.

\subsection{Ice samples}
The behavior of adsorbed CO$_2$ is studied on four different solid water substrates, of which one is crystalline and the other three are amorphous. The crystalline sample is an hexagonal ice crystal with an ordered dangling-proton structure on the surface. This substrate has well-defined binding sites and allows to study the adsorption of CO$_2$, with minimal influence of local surface inhomogeneities. Furthermore, the crystalline structure allows to generate relatively small samples, to study with DFT. The amorphous ice substrates are of direct astrophysical interest and the results can be compared to various laboratory experiments. 

\textit{Hexagonal ice:}\\ Hexagonal ice is a proton-disordered crystal. The oxygen atoms occupy tetrahedrally coordinated lattice positions but the hydrogen atoms form a random hydrogen bond network, according to the Bernal-Fowler rules~\cite{Bernal1933}. At the surface of the basal plane of ice Ih, these rules cannot be satisfied, leaving a random pattern of hydrogen atoms sticking out of the surface, which do not form a hydrogen bond. This dangling bond pattern has a strong effect on the surface energy~\cite{Buch2008,Pan2008} and on the energetics of adsorbed molecules~\cite{Sun2012,Batista2001,Karssemeijer2012}. The lowest energy surface of ice Ih is believed to be the so-called Fletcher phase~\cite{Fletcher1992}, where the dangling protons are ordered, aligned in rows on the surface. In this work, we have used this surface-ordered phase to minimize effects of the disordered proton structure.

To generate the sample, a 144 molecule unit cell of bulk hexagonal ice, with negligible net dipole moment and with the experimental $c/a$ ratio~\cite{Rottger1994}, was generated using the method proposed by~\citet{Buch1998}. The water molecules are arranged in three bilayers, with the Fletcher phase proton order imposed upon the interface between two of the bilayers. Later in the procedure, this interface is used to create the Fletcher phase dangling bond pattern on the surface. The bulk initial sample was optimized with both DFT and TIP4P/2005f by subsequently relaxing the cell volume, freezing the positions of the molecules in the lowest bilayer, adding vacuum (11~\AA~for the DFT sample and 100~\AA~for the pair potential system) along the $z$-axis (parallel to the $c$-axis of the crystal) to create a surface, and finally relaxing the coordinates of all free atoms in the system. To respect the 10~\AA~interaction cutoff of the PP interactions, all calculations with this method were performed on a sample which is duplicated along the $x$ and $y$ directions (parallel to the surface). The densities of the samples are 0.96 and 0.94 g~cm$^{-3}$ for the DFT and PP substrates respectively. \new{The base area of the simulation box is $31\times27$~\AA$^2$~for the PP substrate.}

\textit{Amorphous ice:}\\ The amorphous ice substrates used in this work are the same as the ones we used to study the dynamics of CO on ASW. For a detailed description of the substrate morphology and the creation procedure, the reader is therefore referred to \citet{Karssemeijer2014}. Three different amorphous ice samples are used. Each of these contains 480 water molecules \new{and has a base area of $25\times 25$~\AA$^2$}. The samples have a density of $1.01\pm0.3$~g~cm$^{-3}$ and an effective surface area of $807\pm7$~\AA$^2$, comparable to that of the PP ice Ih sample (838~\AA$^2$).

\section{Results \& Discussion}
\label{sec:results}
\subsection{H$_2$O-CO$_2$ complex}
\label{sec:resultsh2ococomplex}
\begin{figure}[h]
\centering
\includegraphics{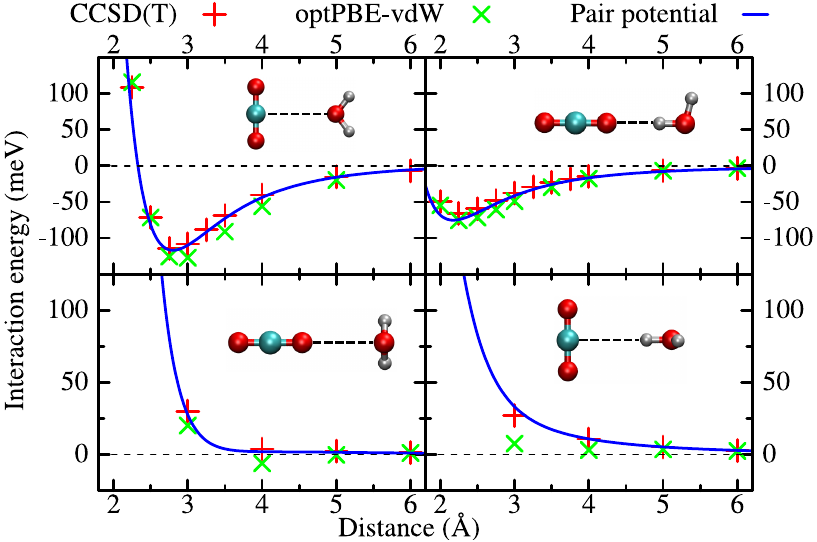}
\caption{\new{Potential energy curve of the H$_2$O-CO$_2$ dimer through four cuts of the potential energy surface at the various levels of theory. The global, $C_{2v}$, minimum is shown in the top left panel.}}
\label{fig:dimer_energy}
\end{figure}

To evaluate the accuracy of the proposed H$_2$O-CO$_2$ pair potential and the optPBE-vdW functional, we shortly evaluate the interaction energy of the complex. \new{In Fig.~\ref{fig:dimer_energy}, we show the interaction energy along four cuts through the potential energy surface (note that this is but a selection of the 30 distinct cuts used to fit the PP). From these curves, we observe that both the pair potential and the DFT calculations reproduce the CCSD(T) points to satisfactory accuracy, albeit that the DFT calculations tend to overbind a bit with respect to the CCSD(T) points}. The top left panel shows the cut through the global minimum, which has a T-shaped, $C_{2v}$, structure with a C-O separation of 2.77~\AA\cite{Makarewicz2010}. Geometry optimizations around the global minimum yield a C-O distance of 2.80~\AA\ and interaction energy of $-117$~meV for the pair potential. DFT geometry optimizations predict a 13~\% stronger interaction energy of $-132$~meV, at a distance of 2.83~\AA.

\subsection{Adsorbed CO$_2$ at low coverage}
\label{sec:tip4pbe}

In this section, the adsorption of CO$_2$ molecules is studied with the PP interactions at the lowest coverages, i.e., one or two adsorbed CO$_2$ molecules per substrate. First, we study the binding sites for a single CO$_2$ molecule on both substrate types and evaluate their binding energies. These results can be compared to submonolayer TPD results. To address the question of CO$_2$ wetting, we also study adsorption energies on crystalline ice when two molecules are present on the substrate to evaluate the strength of the CO$_2$-CO$_2$ interactions.

The binding sites for a single CO$_2$ molecule are found with the adaptive Kinetic Monte Carlo (AKMC) technique~\cite{Henkelman2001,Karssemeijer2012} on both types of ice substrate. Although this is in principle a dynamical simulation technique, it is also very suitable for exploring minima on potential energy surfaces. With the binding sites known, we calculate their binding energies and evaluate their distributions.

\begin{figure}[h]
\centering
\includegraphics{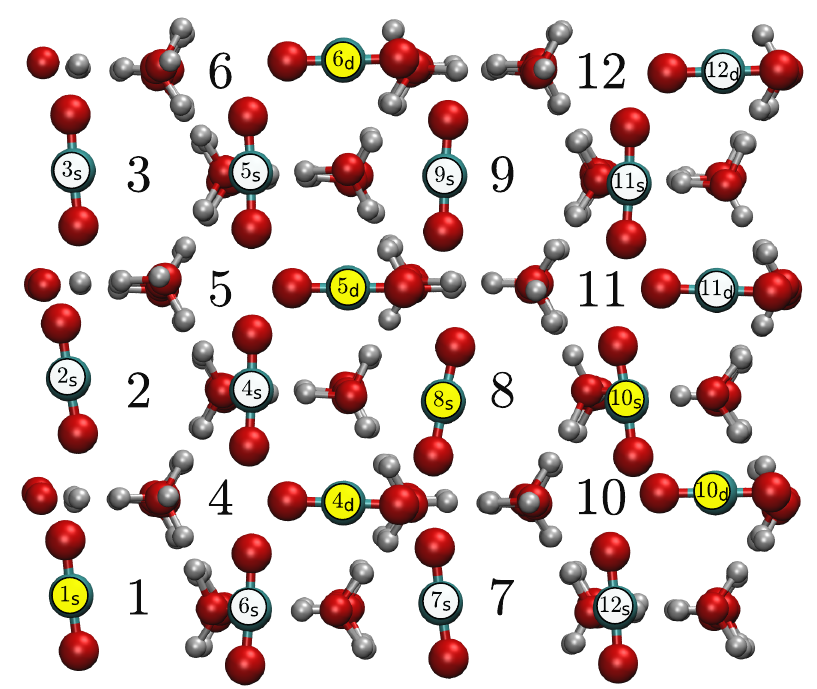}
\caption{Binding sites of CO$_2$ on the hexagonal ice substrate, as found by the pair potential model, are divided into deep (subscript $d$) and shallow (subscript $s$) sites. Sites in yellow are also evaluated at the DFT level.}
\label{fig:1h_bindingsites}
\end{figure}

\begin{figure}[h]
\centering
\includegraphics{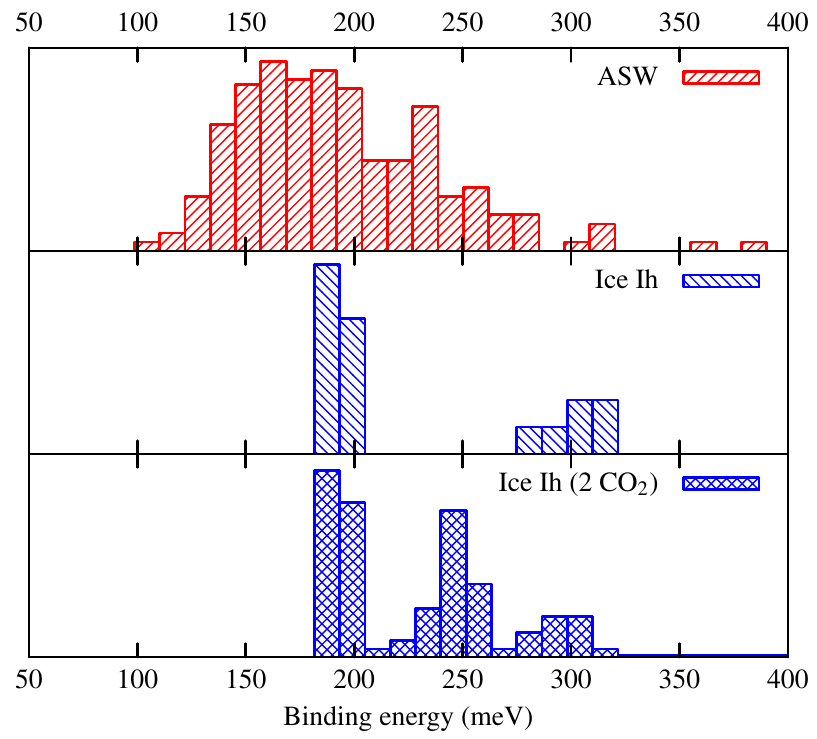}
\caption{Pair potential binding energies, for single CO$_2$ binding sites on the amorphous surfaces and Fletcher's striped phase of hexagonal ice. The bottom panel shows the binding energy distribution on hexagonal ice per CO$_2$ molecule, when 2 CO$_2$ molecules are adsorbed.}
\label{fig:be_hist}
\end{figure}

\begin{table}[h]
\centering
\caption{Binding energies (in meV), and DFT dipole moment (in D), for single CO$_2$ binding sites on the ice Ih substrate.}
\begin{tabular}{cccc}
\hline
Site & B.E., PP & B.E., DFT & Dip.\\\hline
$4_\text{d}$ & 309 & 392    & 0.23 \\
$5_\text{d}$ & 288 & 360    & 0.23 \\
$6_\text{d}$ & 316 & 358    & 0.25 \\
$10_\text{d}$ & 285 & 351   & 0.27\\
$1_\text{s}$ & 194 & 258    & 0.44\\
$8_\text{s}$ & 204 & 273    & 0.46\\
$10_\text{s}$ & 185 & 263   & 0.48\\
\hline
\end{tabular}
\label{tab:singleco2}
\end{table}

\textit{Hexagonal ice:}\\ On the ice Ih sample, 18 unique binding sites are discovered for a single admolecule. Due to the ordered dangling proton pattern on the surface of the Fletcher phase, the adsorption sites are limited in number and well defined. The binding sites are shown in Fig.~\ref{fig:1h_bindingsites} and the distribution of their binding energies is shown in Fig.~\ref{fig:be_hist}. The sites can be classified according to their geometry and binding energy into two categories: deep and shallow. We have numbered the hexagons on the surface and labeled the sites accordingly, together with their category. The deep sites have binding energies around 300~meV, which derives from the interaction of the CO$_2$ molecule with two dangling protons on the surface. The shallow sites have binding energies around 200~meV. 

The binding energy of the deep sites is higher than the experimental values for CO$_2$ adsorption on crystalline ice (216~meV by \citet{Noble2012a}, 220~meV by \citet{Andersson2004}, and 206~meV by \citet{Galvez2007}). We believe this is partly explained by the Fletcher ordering of the dangling OH bonds, which provides very favorable binding sites. To confirm this, we also performed calculations on three proton disordered hexagonal ice samples~\cite{Karssemeijer2012}, were we find a broader distribution of binding energies, which peaks at a lower value of 230~meV.   

Configurations with two adsorbed CO$_2$ admolecules were also investigated on the ice Ih substrate. Geometry optimizations were performed, starting from configurations with two CO$_2$ molecules, both occupying one of the single-molecule adsorption sites. This led to 93 new, unique configurations\footnote{In theory, there should be 153 configurations, but some of the geometry optimizations led to the same local minimum on the potential energy surface.}, for which we calculated the binding energy per CO$_2$ molecule. This distribution is little different from the distribution of singly occupied sites (see Fig.~\ref{fig:be_hist}). The only new feature is a third peak, around 250~meV, which corresponds to states where one deep, and one shallow site is occupied. 

To investigate the effect of the CO$_2$-CO$_2$ interactions on the binding energy, we compared the binding energy per molecule between doubly occupied state and the two corresponding singly occupied states (this was only done for those configurations were the orientations of the two CO$_2$ molecules did not differ from those in their respective singly occupied sites). With this approach, we found only negligible interactions for the majority of configurations. Only three exceptions were found, when two deep sites are occupied, with the two molecules aligned along their axes (configurations $4_d$-$10_d$, $5_d$-$11_d$, and $6_d$-$12_d$). Here we observe a repulsive interaction around 50~meV. Judging by the configurations though, this reduced binding energy arises because in this configuration, the dangling H atoms cannot accommodate both CO$_2$ molecules at the same time so this is rather an effect from the substrate, than a direct CO$_2$-CO$_2$ repulsion. For the remaining configurations, the CO$_2$-CO$_2$ interaction is 0~meV on average, and always between -20 and 20~meV. No significant dependence on CO$_2$-CO$_2$ distance, or their relative orientation was found.

The above suggests that there is little influence from the CO$_2$-CO$_2$ interactions on the binding energy. Hence, based on this analysis, we do not expect CO$_2$ to form islands on crystalline ice substrates. In the PP model however, the effects from many-body interactions are intrinsically missing because it was fitted to data from the gas-phase complex. These effects could still have a big influence and so, to shed some more light on this aspect, several of the singly and doubly occupied states are further investigated at the DFT level in Section~\ref{sec:dft}.

\textit{Amorphous ice:}\\ On the amorphous substrates, a total of 365 states are discovered by the AKMC simulations. For these simulations, the water molecules in the substrates are kept frozen in order to avoid significant changes in the water substrate. This typically results in about 10~\% lower binding energies than on the unconstrained substrates~\cite{Karssemeijer2014}, because the substrate cannot fully accommodate the guest molecule. The distribution of binding energies on the amorphous substrates is broadly peaked around 190~meV. This is in good agreement with the submonolayer values of 202~meV on non-porous ASW from~\citet{Noble2012a} and $181\pm78$~meV by~\citet{Edridge2010} and~\citet{Edridge2013} , who also report a broad distribution of binding energies.  

\subsection{DFT calculations \& many-body effects}
\begin{table*}[hbtp]
\centering
\caption{Binding energies (meV), per molecule, of CO$_2$ to specific binding sites on the ice Ih substrate for double occupancy. The extracted CO$_2$-CO$_2$ interaction energy (meV), CO$_2$ molecular dipole moments (D), and dipole-dipole interactions (meV) from the DFT structures are also listed.}
\begin{tabular}{ccccccccc}
\hline
Site 1 & Site 2 & B.E., PP & B.E., DFT & E$_{\text{CO$_2$-CO$_2$}}$, PP & E$_{\text{CO$_2$-CO$_2$}}$, DFT  & Dip.$_1$ & Dip.$_2$ & Dip.-dip. int.\\\hline
$4_\text{d}$ & $5_\text{d}$ & 	296 & 	376 & 5 & 0 & 0.21 & 0.22  & 0.4 \\
$4_\text{d}$ & $6_\text{d}$ & 309 & 375 & 6 & 0  & 0.22 & 0.24 & 0.5\\
$4_\text{d}$ & $10_\text{d}$ & 273 & 363 & 47 & 15 & 0.22 & 0.26& 0.0 \\
$4_\text{d}$ & $1_\text{s}$ & 257  & 329 & -11 & -7  & 0.22 & 0.49 & 0.3\\
$5_\text{d}$ & $6_\text{d}$ & 294 & 355 & 15 & 8 & 0.23 & 0.25 & 0.4\\
\hline
\end{tabular}
\label{tab:doubleco2}
\end{table*}

\label{sec:dft}

\new{Adsorbate-substrate interactions are much more complicated than gas phase dimer interactions and it is important to make sure that no essential features are missing in the PP model. For this reason, we have performed DFT calculations to explicitly evaluate the contribution of induced many-body electrostatic interactions, which is the crucial missing ingredient in a PP model, when one uses gas-phase reference data. One has to keep in mind though, that the adsorption of small molecules is one of the biggest challenges in DFT, even with modern methods to account for electron dispersion~\cite{Klimes2012}. Indeed, calculations of the adsorption energies of small molecules to graphene and benzene also show serious overbinding within the vdW-DF formalism, while the equilibrium geometries are typically somewhat better predicted~\cite{Hamada2012,Silvestrelli2014}. The DFT results on the binding energies should therefore not a priori be seen as more reliable than the PP results. In view of the CO$_2$ wettability on water substrates, we need to evaluate the dipole moments in the adsorbed CO$_2$ molecules, which are induced by the water molecules. These induced moments could lead to an additional attraction between the adsorbates, possibly leading to a non-wetting behavior.}

To evaluate these interactions, several of the singly and doubly occupied sites on the hexagonal ice substrate, as described in Section~\ref{sec:tip4pbe}, are studied at the DFT level. Starting from the DFT optimized water substrate, with the CO$_2$ molecule(s) positioned at the PP positions, geometry optimizations were performed for seven single (see Fig.~\ref{fig:1h_bindingsites}), and five doubly occupied states. During these relaxations, little change was observed in the atomic coordinates (0.5~\AA\ at the maximum), suggesting that the local minima found with the PP are in good agreement with those of the DFT calculations. 

For each configuration, the binding energy per CO$_2$ molecule was calculated with respect to calculations on the isolated ice substrate and CO$_2$ monomer. These are reported in Table~\ref{tab:singleco2} for the single binding sites, and in Table~\ref{tab:doubleco2} for the doubly occupied configurations. The binding energies are significantly stronger in the DFT case, than those from the PP calculations. Qualitatively, the weaker bonding of the shallow sites, compared to the deep ones, is well reproduced by the DFT calculations, but the energetic ordering for sites of equal type is not the same. On the quantitative level, DFT is seen to overbond by 10 to 40\% with respect to the PP results, for the singly occupied sites. This is more than expected, based on the difference of 13\% between the two methods found from the calculations on the gas phase complex. \new{In the presence of a substrate however, interactions do not all arise from the attractive regions in the potential energy surface, but also from the repulsive parts. Based on the data in Fig.~\ref{fig:dimer_energy}, the DFT calculations also overbind (with respect to the \textit{ab-initio} data) in this region, which further contributes to the strong binding energies.}

Analogous to the PP calculations, we evaluated the CO$_2$-CO$_2$ interactions for the double occupied sites by comparison of the total binding energy with respect to the single molecule binding energies. As can be seen from Table~\ref{tab:doubleco2}, this effect is only marginal. With respect to the PP calculations, the CO$_2$-CO$_2$ interactions are smaller for the DFT calculations, but they have the same sign. The strongest effect, just like in the PP case, is the repulsion when two CO$_2$ molecules occupy two deep sites which are aligned with each other (site 4$_d$-10$_d$).

Using the MLWF approach, the dipole moments, induced by the water substrate in each adsorbed CO$_2$ molecule, are calculated. With these dipole moments, it is then possible to make an estimate of the interaction between the CO$_2$ molecules, due to polarization by the substrate. The induced dipole moments are listed in Tables~\ref{tab:singleco2} and~\ref{tab:doubleco2}. The magnitude of the induced moments depends strongly on the type of adsorption site. Deep sites have relatively small moments of about 0.25~D, while the shallow sites have almost twice as large dipole moments. This large variation of dipole moments, was also found for adsorption of water monomers by~\citet{Sun2012} and is explained by the large variations in the local electric field on the water ice surface due to the dangling OH bonds~\cite{Pan2008,Watkins2011}. A comparison of the CO$_2$ dipole moments at a specific position, with and without the presence of a second CO$_2$ molecule shows that this presence has little to no effect on the induced dipole (remember that this dipole is induced primarily by the water substrate). The calculated dipole-dipole interactions for the doubly occupied sites (see Table~\ref{tab:doubleco2}) are all less than one meV, much smaller than the spread in the interaction energies due to the H$_2$O-CO$_2$ interactions. Even when assuming parallel alignment of two of the strongest induced dipoles that we find, $\sim$0.5~D, at the smallest distance, about 5~\AA, the dipole-dipole interaction energy is only 1.2~meV. Thus, the induced electric moments do not lead to a sufficiently large additional attraction between the CO$_2$ molecules to change the adsorption behavior from wetting to non-wetting.

\subsection{Ballistic deposition simulations}

\begin{figure*}[htbp]
\centering
\includegraphics{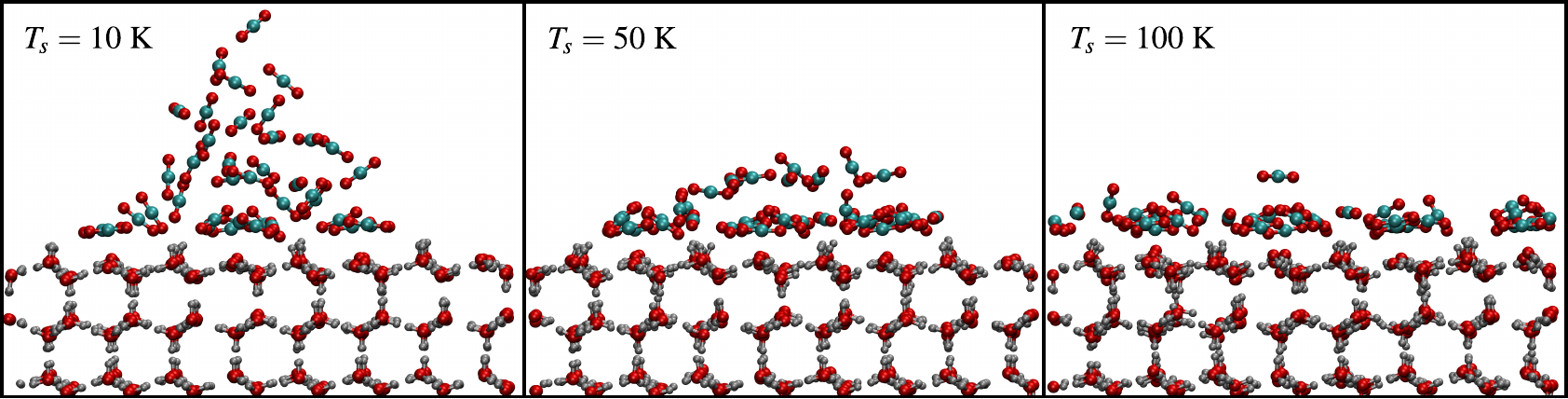}
\caption{Typical structures obtained after depositing 40 CO$_2$ molecules from the center of the cell with a gas temperature of 300~K at substrate temperatures of 10, 50, and 10~K.}
\label{fig:md_structures}
\end{figure*}

To simulate the behavior of CO$_2$ on water ice surfaces under laboratory conditions, molecular dynamics (MD) simulations were performed to simulate the deposition of CO$_2$ onto two of the solid water substrates: the crystalline sample and one of the three amorphous substrates. Again, we were specifically interested in whether or not we could observe CO$_2$ island formation. In these simulations, the substrate is equilibrated at a substrate temperature, $T_{s}$, and is bombarded with gas phase CO$_2$ from either a single point at the $x,y$-center of the box, or from a random point above the substrate, well outside the potential cutoff radius. Although the deposition from a random point is clearly more representative of the laboratory situation, we also included the depositions from the center because this is the most favorable situation for island formation. The incident CO$_2$ molecules are given a random orientation and their velocity is drawn from a Maxwell-Boltzmann distribution of temperature $T_{g}$.  The direction of the velocity of the incident molecules is also random, but such that the angle with the surface normal (the $z$-axis) is less than 10$^{\circ}$. Experimentally, this corresponds to direct a deposition at 90$^{\circ}$, not to background dosing. The molecules have no rotational or vibrational energy. 

The depositions are done one molecule at a time and each deposition consists of two steps. First, the deposition itself is carried out in a run of 37.5~ps in the NVE ensemble. This is sufficiently long for the incident CO$_2$ molecule to reach the surface and dissipate its kinetic energy into the substrate. In the second step, the additional energy which was added to the system is removed by re-equilibrating the system (which now has one extra CO$_2$ molecule) for 50~ps in the NVT ensemble using a Nos\'e-Hoover thermostat at temperature $T_{s}$. This procedure is reminiscent of a cryostat's function in the laboratory. After this equilibration, the next CO$_2$ molecule is deposited and the whole procedure is repeated until 40~CO$_2$ molecules, about one monolayer, have been deposited. The gas temperature, $T_{g}$, is either 300 or 50~K and the substrate temperature, $T_{s}$ is either 100, 50, or 10~K. For each of the six temperature combinations, the whole deposition simulation was carried out at least four times to gather statistics.

Even though these deposition simulations are carried out in order to mimic the experimental deposition of a CO$_2$ layer, the timescales are very different. In our simulations, the incoming CO$_2$ flux is on the order of $10^{23}$ molec~cm$^{-2}$s$^{-1}$, while in experiments, the flux is on the order of $10^{12-15}$ molec~cm$^{-2}$s$^{-1}$. 

\new{To quantify the wetting behavior of an adsorbate, one typically considers the contact angle. However, because our deposited structures are on the nanoscale, we cannot determine this quantity accurately and we introduce two different quantities to analyze the structure instead}. The first, $\Delta z$, relates to the height of the deposited CO$_2$ structure. It is defined as the difference in $z$-coordinate between the highest and lowest CO$_2$ molecule (based on the $z$-coordinate of their respective centers of masses). The second quantity, $\Delta R$, describes the spatial extent of the CO$_2$ structure in the directions parallel to the surface. It is defined as $\Delta R = \langle \sqrt{ (\bm{r}\cdot\hat{\bm{x}})^2 + (\bm{r}\cdot\hat{\bm{y}})^2} \rangle$, where the average is over all CO$_2$ molecules and $\bm{r}$ are the CO$_2$ center of mass coordinates, measured from the center of mass of the entire CO$_2$ distribution. The vectors with hats indicate unit vectors along the respective Euclidean axes.

\textit{Results:}

Visual analysis of the structures, after 40~CO$_2$ have been deposited, reveals immediately that no islands are formed when the deposition is done from a random position. When the deposition is done from the center of the box, islands, or rather, towers are formed when the substrate temperature is either 50 or 10~K. At 100~K however, the CO$_2$ molecules already fully wet the ice surface, even at these short timescales. This behavior is illustrated in Fig.~\ref{fig:md_structures}, where typical deposited structures are shown after deposition of 40~CO$_2$ molecules on the hexagonal ice sample, from the center of the box, with a gas temperature of 300~K. \new{Judging from the obtained structures, the contact angle of the islands formed at $T_s=10$~K, is between 0 and 90$^\circ$, indicating a high wettability.} Furthermore, it is interesting to note that we observed three events where a CO$_2$ molecule desorbed from the surface. This happened at the highest substrate temperatures of 50 and 100~K. Due to the low occurrence of these events though, we performed no further investigations.

\begin{figure}[h]
\centering
\includegraphics{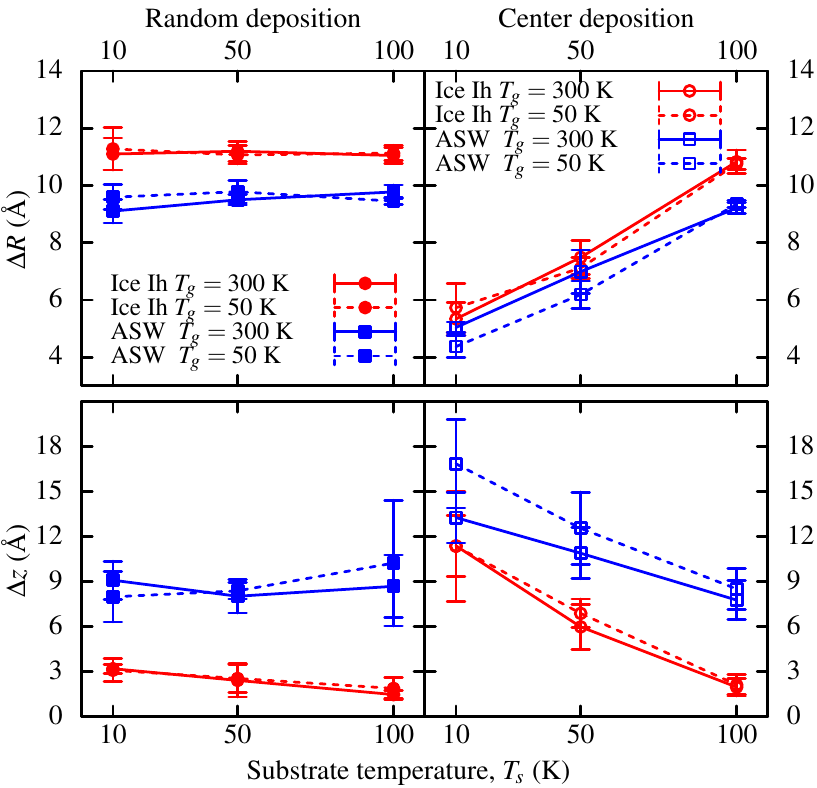}
\caption{Horizontal and vertical structure of the CO$_2$ layer, after 40 deposited molecules, described by the $\Delta z$ and $\Delta R$ quantities.}
\label{fig:comr_deltaz}
\end{figure}

The results on the two structural quantities defined above are shown in Fig.~\ref{fig:comr_deltaz}. Here, the uncertainties arise from averaging over the repeated simulations. From the behavior of both quantities, it is clear that the largest influence on the resulting CO$_2$ structure comes from the substrate temperature, $T_s$, and the deposition method (from the center, or from a random point above the substrate), while the gas temperature, $T_g$, has almost no influence. 

When the deposition is artificially constrained to originate from the center of the box, the tower-like structures can be seen from the behavior of $\Delta z$. At substrate temperatures of 10~K, $\Delta z$ is significantly higher for this deposition method than in the case of the depositions from a random position on both the ice Ih and the ASW substrate. As the substrate temperature increases, the difference between the deposition methods vanishes due to the collapse of the CO$_2$ structure, forming a uniform layer on the ice surface. The same behavior is observed for $\Delta R$, which increases with increasing $T_s$ due to the flattening of the CO$_2$ structure. 

The difference between the ASW and ice Ih substrates is best seen from $\Delta z$ at high substrate temperatures. On the crystalline substrate, the CO$_2$ molecules fully wet and form a monolayer (approximately 3~\AA\ thick) while on the ASW substrates, the CO$_2$ molecules are able to migrate into the nanopores of the substrate. This is why the $\Delta z$ values are consistently higher on ASW than on the ice Ih. The opposite holds for $\Delta R$, but this is due to the slightly larger base area of the simulation box of the crystalline sample, which results in larger maximal values of $\Delta R$ (when the adsorbate fully wets the surface).

\begin{figure}[h]
\centering
\includegraphics{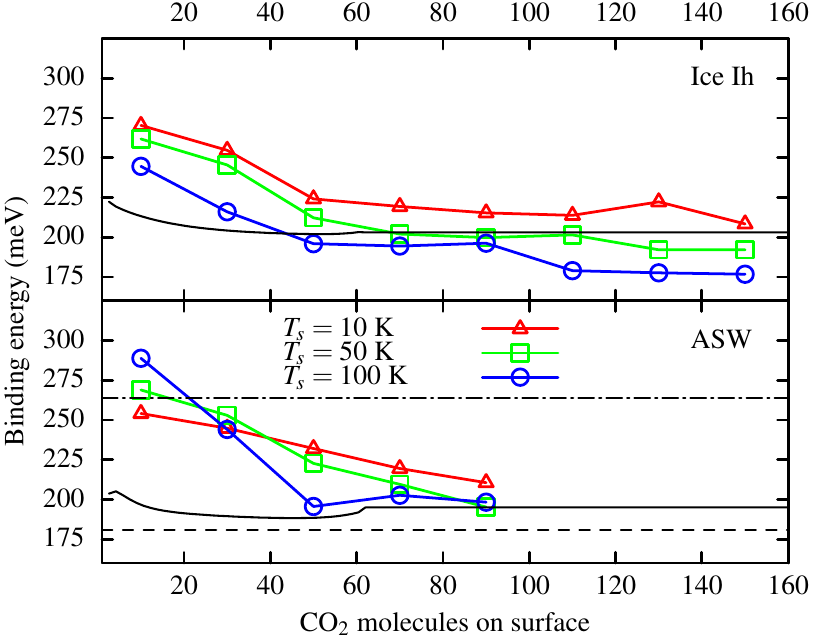}
\caption{CO$_2$ binding energy as a function of surface coverage, extracted from the MD deposition simulations on ice Ih (top panel) and ASW (bottom panel). The solid black lines are the coverage-dependent functions from experiments by \citet{Noble2012a} on ASW and crystalline ice respectively. The monolayer (dashed) and multilayer (dash-dotted) desorption energies by \citet{Edridge2013} for CO$_2$ from ASW are also shown.}
\label{fig:be_mddep}
\end{figure}

Besides the atomic structure, also the energetics of the ballistic deposition simulations are interesting. In particular, they can be used to extract the CO$_2$ binding energy, as a function of surface coverage, which is the quantity probed by TPD experiments. We analyzed these binding energies from our simulations which most closely resemble the experimental situation: the depositions from a random position. To cover the full transition from the submonolayer to the multilayer regime we continued the deposition simulations until at least 100~CO$_2$ molecules ($\sim$~2.5 monolayers) were deposited on the surface. 

In Fig.~\ref{fig:be_mddep} we show the binding energy of CO$_2$ as a function of surface coverage for the ASW and ice Ih substrates at the three different substrate temperatures. 
 The binding energies were obtained from the decrease of the total potential energy of the configurations after each NVT-equilibration run, as the number of CO$_2$ molecules on the surface increases. To average out the fluctuations as much as possible, we averaged the binding energies over sets of 20 deposited molecules and, since we observe no quantitative difference, also over the two gas temperatures.  

As seen from Fig~\ref{fig:be_mddep}, the adsorption energy of CO$_2$ on both amorphous and crystalline water ice is stronger at submonolayer coverage than in the multilayer regime. This shows that the CO$_2$-H$_2$O interaction is stronger than the CO$_2$-CO$_2$ interactions, which again, leads to the expectation that a CO$_2$ layer will show wetting behavior on a solid water substrate. On both substrates, a constant multilayer desorption energy is reached when about 50 molecules, or just over one monolayer, have been deposited. To check convergence, we went up to 160 deposited molecules on the ice Ih surface. Averaged over the molecules in this multilayer regime, we find a desorption energy of $224\pm16$~meV on crystalline ice and $230\pm15$~meV on ASW. These values are close to the multilayer values of 195 and 203~meV by \citet{Noble2012a} on non-porous ASW and crystalline H$_2$O respectively, but significantly lower than the $264\pm15$~meV reported by~\citet{Edridge2013}. 

At the lowest coverage we find similar binding energies on both substrate types: around 260 meV. This is consistent with the results of the static calculations (Fig.~\ref{fig:be_hist}). An interesting difference between the substrates is observed in the temperature dependence at low coverage. On ice Ih, we find the highest binding energies at the lowest temperatures. This is the expected behavior because at low temperature, the thermal contributions are the smallest. On ASW however, we observe just the opposite trend. Here the binding energies increase with temperature. Although the uncertainties are large, one can speculate that this counterintuitive behavior is due to the morphology of the ASW substrate, which has surface nanopores with strong binding energy~\cite{Karssemeijer2014}, which may only be reached on these timescales if sufficient thermal energy is available.

\new{Finally, to assess the stability of the CO$_2$ towers formed in the simulations at $T_s=10$~K when deposition originates from the center of the box, we performed additional MD runs were we heated the resulting structures during 200~ps, up to a temperature of 100~K. We observed that the structures start to collapse at around 40~K, to form a fully covering overlayer by the end of the simulation. This shows that these artificial towers are not equilibrium structures and we do not expect to see them on experimental timescales.}


\section{Conclusions}
The adsorption behavior of CO$_2$ on amorphous solid water and the proton ordered, Fletcher phase of ice Ih has been studied using a new pair potential for H$_2$O-CO$_2$ interactions. This potential was fitted to \textit{ab-initio}, CCSD(T) calculations on the gas phase H$_2$O:CO$_2$ complex and was subsequently used in static calculations on CO$_2$ adsorption at low coverages and in molecular dynamics simulations of the deposition of CO$_2$ layers on both kinds of ice surfaces.  Following the recent experimental claim of CO$_2$ being a non-wetting molecule on water ices, we paid specific attention to this aspect throughout the paper. 

In the gas phase, the binding energy of the CO$_2$:H$_2$O complex is about twice as strong as that of the CO$_2$:CO$_2$ dimer. Hence, one would expect CO$_2$ to show full wetting behavior on a H$_2$O surface, instead of forming islands. This is also the main conclusion from our work. However, as experiments suggest, the difference in strength between the CO$_2$-H$_2$O and the CO$_2$-CO$_2$ interactions is much smaller in the presence of a water surface, than in the gas phase. This can be understood if one considers that the hydrogen bonds between the water molecules are much stronger still. This means that the water molecules will not easily accommodate the CO$_2$ adsorbants, leading to an effective weakening of the CO$_2$-H$_2$O interactions.

From our study into the binding sites of CO$_2$ on hexagonal ice, we find  that at low coverage, CO$_2$ molecules have no preference towards occupying binding sites which are spatially close together, which would lead to the onset of island formation. From additional DFT calculations on these structures, we showed that induced electrostatic interactions, which are not included in the pair potential model, do not alter this behavior. 

From dynamical simulations of the deposition of CO$_2$ molecules on both types of ice substrates we arrive at the same conclusion. Under typical laboratory conditions, in terms of substrate temperature and deposition method, we observe no formation of CO$_2$ islands on either the crystalline or the amorphous ice substrates. Clustering of the CO$_2$ molecules was only observed using a rather unphysical deposition method (always from the same point), when the substrate temperature is very low (10~K).

With increasing  CO$_2$ coverage, we observe a decreasing binding energy. In the multilayer regime, our results are in good agreement with the TPD experiments by \citet{Noble2012a}, but we don not see any hints of island formation, as derived from the experimental results. Because the complicated double-peaked TPD spectra, on which this conclusion was based, is only observed on ASW, and not on crystalline ice, we rather believe these spectra to result from the substrate morphology, and not from the intrinsic wetting behavior of CO$_2$ on water surfaces. \new{Especially small pores on the substrate surface may play an important role. Even though the ice is characterized as non-porous, small pores could still be present and may provide favorable binding pockets which can lead to the complicated TPD spectra. In our MD simulations, we indeed see hints of CO$_2$ molecules moving into nanopores on the surface. To quantify this however, follow-up investigations into the long-timescale, thermal behavior of H$_2$O:CO$_2$ systems are needed.} Regarding the energetics of the wetting behavior, the experimental difference in binding energy between the submonolayer and multilayer regime, which should lead to island formation, is less than 10~meV. This is much less than the binding energy differences which we observe from site to site, arising from the local structure of the water ice. We therefore believe this is insufficient to lead to the formation of islands. Finally, in similar TPD experiments of CO$_2$ on ASW by \citeauthor{Edridge2013}\cite{Edridge2010,Edridge2013}, the complex TPD peaks are not observed, albeit they arrive at very different desorption energies.

With this work we have presented, and tested, a set of intermolecular forcefields for the atomistic modeling of mixed H$_2$O:CO$_2$ systems. The forcefields are computationally efficient and can be used in future research to study more complex processes in the bulk of the ice, to shed more light on the segregation of mixed ices in the interstellar medium.

\section*{Acknowledgements}
This work has been funded by the European Research Council (ERC-2010-StG, Grant Agreement no. 259510-KISMOL), and the VIDI research programme 700.10.427 and the research programme of the "Stichting voor fundamenteel Onderzoek der Materie (FOM)", both financed by The Netherlands Organization for Scientific Research (NWO). We gratefully acknowledge Ad van der Avoird for help with the zero point energy corrections and Stefan Andersson for performing the \textsc{Gaussian03} calculations.

\footnotesize{
\bibliography{paper} 
\bibliographystyle{rsc} 
}

\end{document}